\documentclass[12pt,preprint]{aastex}

\newcommand{\greeksym}[1]{{\usefont{U}{psy}{m}{n}#1}}

\newcommand{\uDelta}{\mbox{\greeksym{D}}}

\newcommand{\etal}{{\em et~al.\/}\ }
\newcommand{\eg}{{\em e.g.\/}\ }

\newcommand{\LOF}{{\em LOF\/}\ }
\newcommand{\ppmlr}{\ensuremath{\mbox{\em ppmlr}}}

\begin{document}

\title{The Numerical Simulation of Radiative Shocks.\\ 
I: The elimination of numerical shock instabilities using a localized oscillation filter}

\author{Ralph S. Sutherland}
\affil{Research School of Astronomy \& Astrophysics, Australian National University}

\author{David K. Bisset\altaffilmark{1}}
\affil{School of Mathematical Sciences and Department of Physics \& Theoretical Physics, Australian
National University}
\and
\author{Geoffrey V. Bicknell}
\affil {Research School of Astronomy \& Astrophysics and Dept. of Physics \& Theoretical Physics,
Australian National University}

\begin{abstract}
We address a numerical instability that arises in the directionally split computation of hydrodynamic flows when
shock fronts are parallel to a grid plane. Transverse oscillations in pressure, density and temperature are
produced that are exacerbated by thermal instability when cooling is present, forming
post--shock `stripes'. These are orthogonal to the classic post--shock 'ringing' fluctuations.  
The resulting post--shock `striping' substantially modifies the flow. We discuss three
different methods to resolve this problem. These include (1) a method based on artificial
viscosity; (2) grid--jittering and (3) a new localized oscillation filter that acts on
specific grid cells in the shock front. These methods are tested using a
radiative wall shock problem with an embedded shear layer. 
	The artificial viscosity method is unsatisfactory since, while it does reduce post--shock
ringing, it does not eliminate the stripes and the excessive shock broadening renders the
calculation of cooling inaccurate, resulting in an incorrect shock location. Grid--jittering
effectively counteracts striping. However, elsewhere on the grid, the shear layer is
unphysically diffused and this is highlighted in an extreme case. The oscillation filter method
removes stripes and permits other high velocity gradient regions of the flow to evolve in a
physically acceptable manner. It also has the advantage of only acting on a small fraction of
the cells in a two or three dimensional simulation and does not significantly
impair performance.
\end{abstract}

\section{Introduction}
\label{s:intro}

Computational hydrodynamics is an established research area in astrophysics and the insight gained by numerical
simulations has advanced our knowledge in many topics including astrophysical jets, accretion discs, cosmology,
star formation, supernovae and stellar structure, to name a few. Most of these simulations have been confined to
the regime of adiabatic flow and in the case of extragalactic jets, for example, this assumption is warranted.
However, when the flow is significantly radiative, as in the case of stellar jets \citep{blondin90a,xu00a}, or when
radiative shocks are important, \eg\ when light jets interact with dense material driving a radiative shock into
the latter, an adiabatic approximation is no longer appropriate. The effect of radiative cooling on the internal
energy needs to be taken into account and this affects the resulting flow substantially. The research presented in
this paper arose from the conduct of such simulations that will be reported in subsequent papers in this series. A
numerical shock instability, which is often present at some level in adiabatic simulations, is exacerbated by
cooling and becomes so severe that it probably invalidates any physical interpretation of the results. This
instability therefore needs to be controlled when conducting simulations containing radiative shocks.

\begin{figure}[tbp]
\plotone{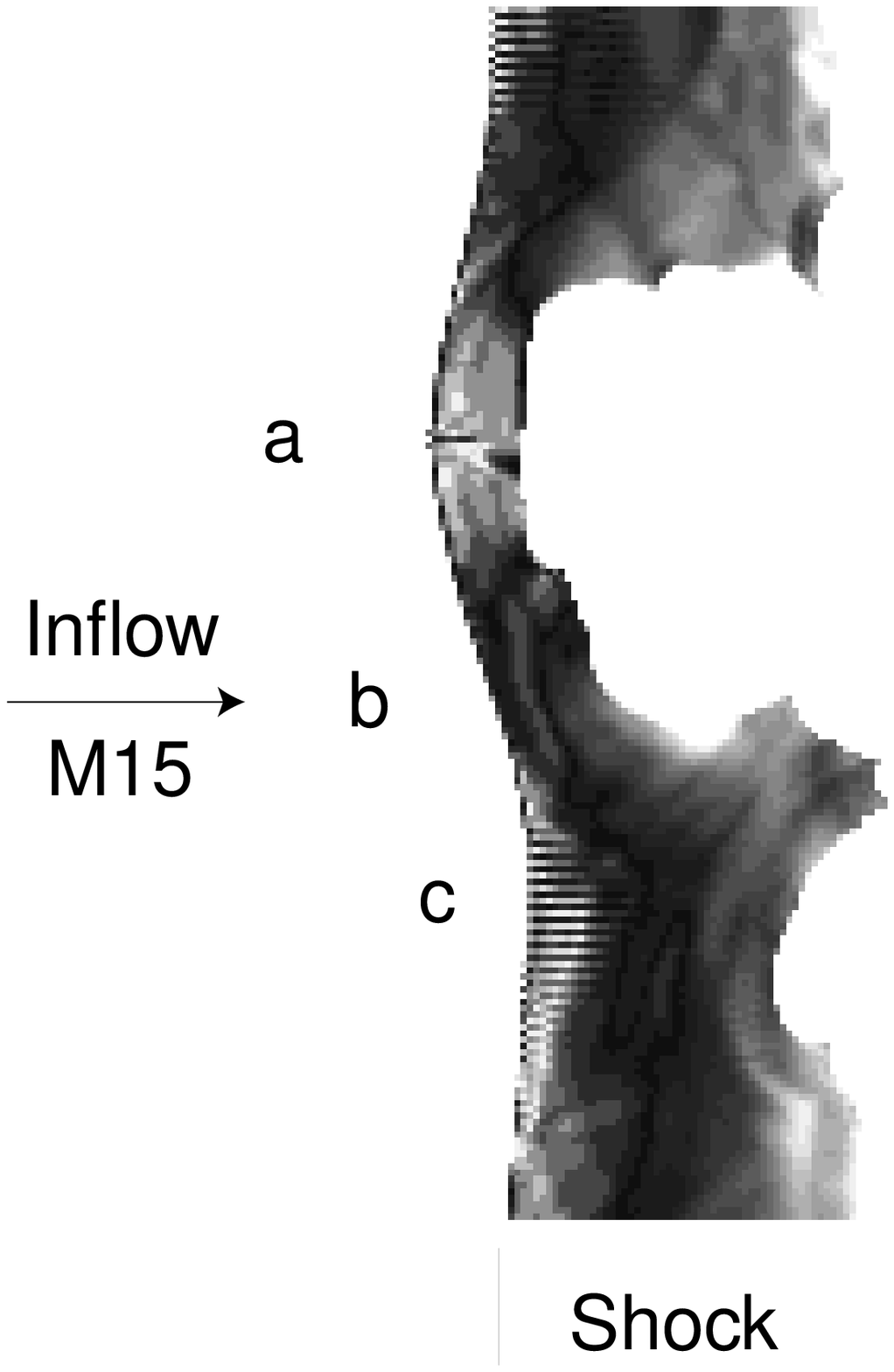}
\caption{}
\label{f:occur}
\end{figure}

The instability to which we are referring is one which occurs in directionally split codes, when a shock propagates
parallel, or nearly parallel, to a coordinate plane, and is most obvious when the shock velocity is small with
respect to the grid (see Figure \ref{f:occur}). Small errors in cells parallel to the shock front are amplified,
leading to trailing perpendicular stripes, representing transverse quasi-periodic density, pressure and temperature
fluctuations, as the shock propagates. In adiabatic flow, this ``striping'' often damps out in the
post--shock region where the sound speed is high, and usually is not a significant cause for concern. However,
when cooling is dynamically important, density fluctuations are thermally unstable; this leads to prominent,
unphysical features in the flow. This problem is not isolated to the code that is the subject of this paper -- a
code based on the {\em VH-1} implementation of the Piecewise Parabolic Method (PPM;
\citet{colella84a}; CW84). The problem is generic and occurs in other directionally split codes. For example,
we have encountered it using both two and three dimensional versions of the {\em ZEUS} code
\citep{norman92a,norman92b}. In part the instability is related to cells that are slightly over-pressured, driving
matter into lower pressure cells.

The source of the striping is described by CW84\footnote{see \S~4 of their paper}. To quote
from their paper: `` In the column of zones where a shock transition occurs, a small disturbance develops in
tangential velocity, due to numerical error. These small velocities transport large amounts of the conserved
quantities, creating in effect, large sources and sinks in the shock transition zone for the essentially
one-dimensional calculations of the motion of the shock in each row of zones. Finally, although we have
discussed these errors in terms of the single-step Eulerian schemes, they occur as well when the PPM
Eulerian scheme is formulated as a Lagrangian step followed by a remap''. As we have mentioned above, this
instability becomes much more important when cooling is introduced. It should also be appreciated that the
seed for the instability may be physical rather than numerical and we give an example of this in
\S~\ref{s:comparison}. For example, a wave propagating along a shock front, may provide the seed for the
instability which is then unphysically amplified by the mechanism indentified by CW84. The fact that
stripes only form when the shock front is almost parallel to one of the grid planes also indicates that the
instability is numerical. 

Standard ways of dealing with post--shock fluctuations include the introduction of either artificial
viscosity or grid ``jittering''. Artificial viscosity smoothes the shock over more than two zones, thereby
reducing the fluctuations between zones parallel to the shock front as it propagates across the grid.
Grid--jittering involves displacement of the grid in an oscillatory fashion thereby smoothing fluctuations
between cells with the additional numerical diffusion from the modified grid resampling process. Both of these
techniques can be useful. However, they have the disadvantage that they are applied to the entire flow. This
means that otherwise sharp features, \eg\ tangential discontinuities in velocity, are smoothed
unnecessarily, degrading the resolution. We give examples of this in
\S~\ref{s:comparison}. We have therefore implemented a new approach involving a ``localized oscillation filter''.
This filter applies a light smoothing to cells parallel to the shock front, in one sweep, using information
stored from the previous orthogonal sweeps. In this case the smoothing is applied to cells local to a
shock \emph{in the previous orthogonal sweep}, even though the current sweep may be parallel to the shock and
should not include any significant velocity variations.  The localized
action of this smoothing avoids many of the undesirable side--effects
of other methods. In the following sections we describe the filter in some detail and then present some
comparison calculations. 

In the work described here we have utilized a version of the {\em VH-1} code, made available by Blondin \etal\
via the web-site (http://wonka.physics.ncsu.edu/pub/VH-1/). We have extensively reorganized the basic code for
vectorization and overall efficiency on the ANU Fujitsu VPP300 computer as well as adding subroutines to advect a
passive scalar that distinguishes different gases and to update the energy density, using an implicit
method, when optically thin radiative cooling cooling operates. The new code is now called, simply, \ppmlr, to
distinguish it from {\em VH-1} and to refer to the method used.

\section{The Local Oscillation Filter}

As mentioned above, methods involving global smoothing in one form or another have been used to suppress
post-shock ringing and stripes.  In this paper, we report a spatially and temporally localized method of
eliminating the striping problem, the `Local Oscillation Filter' (hereafter \LOF),  which
interferes minimally with most of the cells in a multi--dimensional simulation.  The filter consequently
preserves the high resolution nature of modern hydrodynamic methods. In particular, it preserves the excellent
shock-capturing characteristics of PPM. 

\begin{figure}[tbp]
\plotone{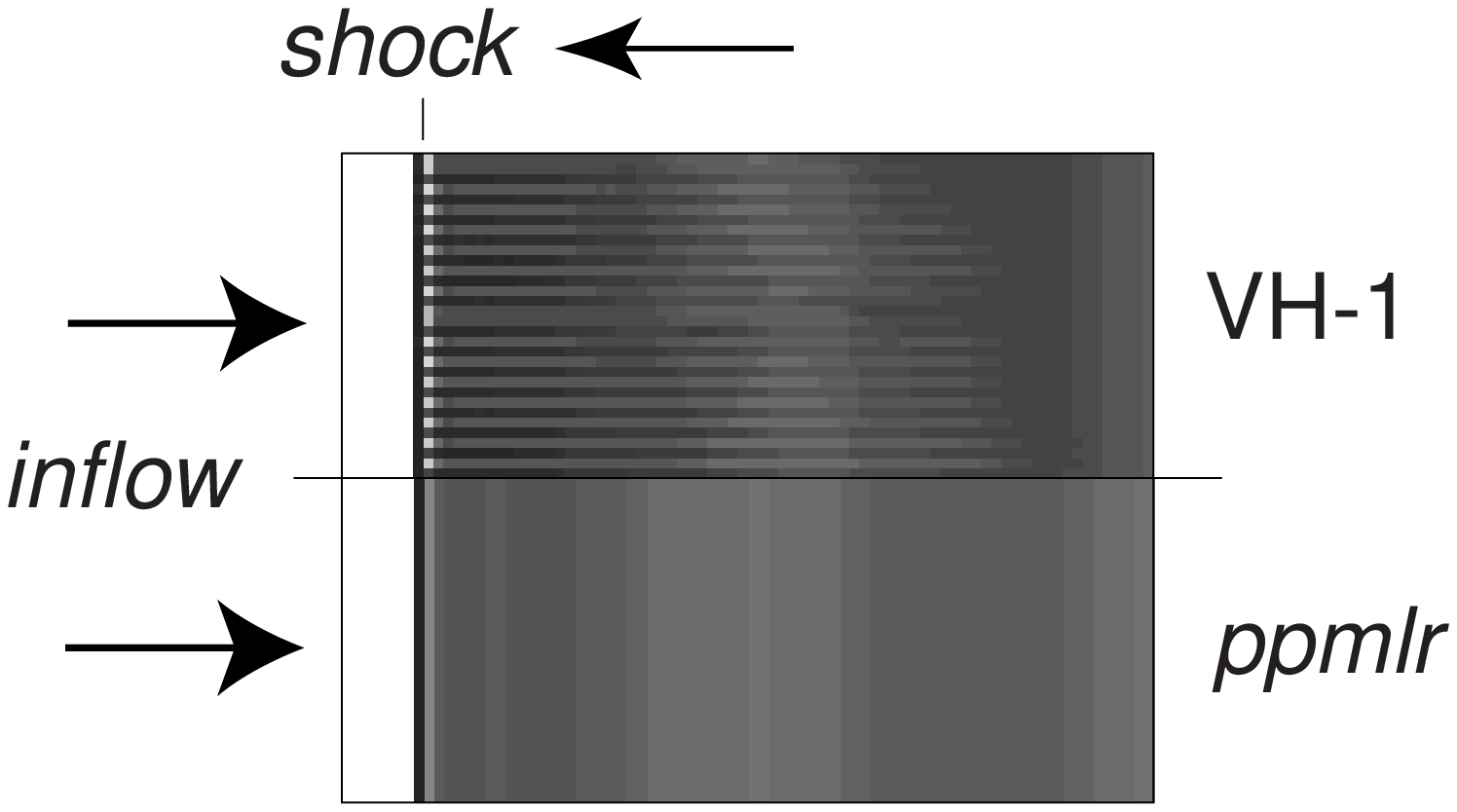}
\caption{}
\label{f:vpstripes}
\end{figure}

\begin{figure}[tbp]
\plotone{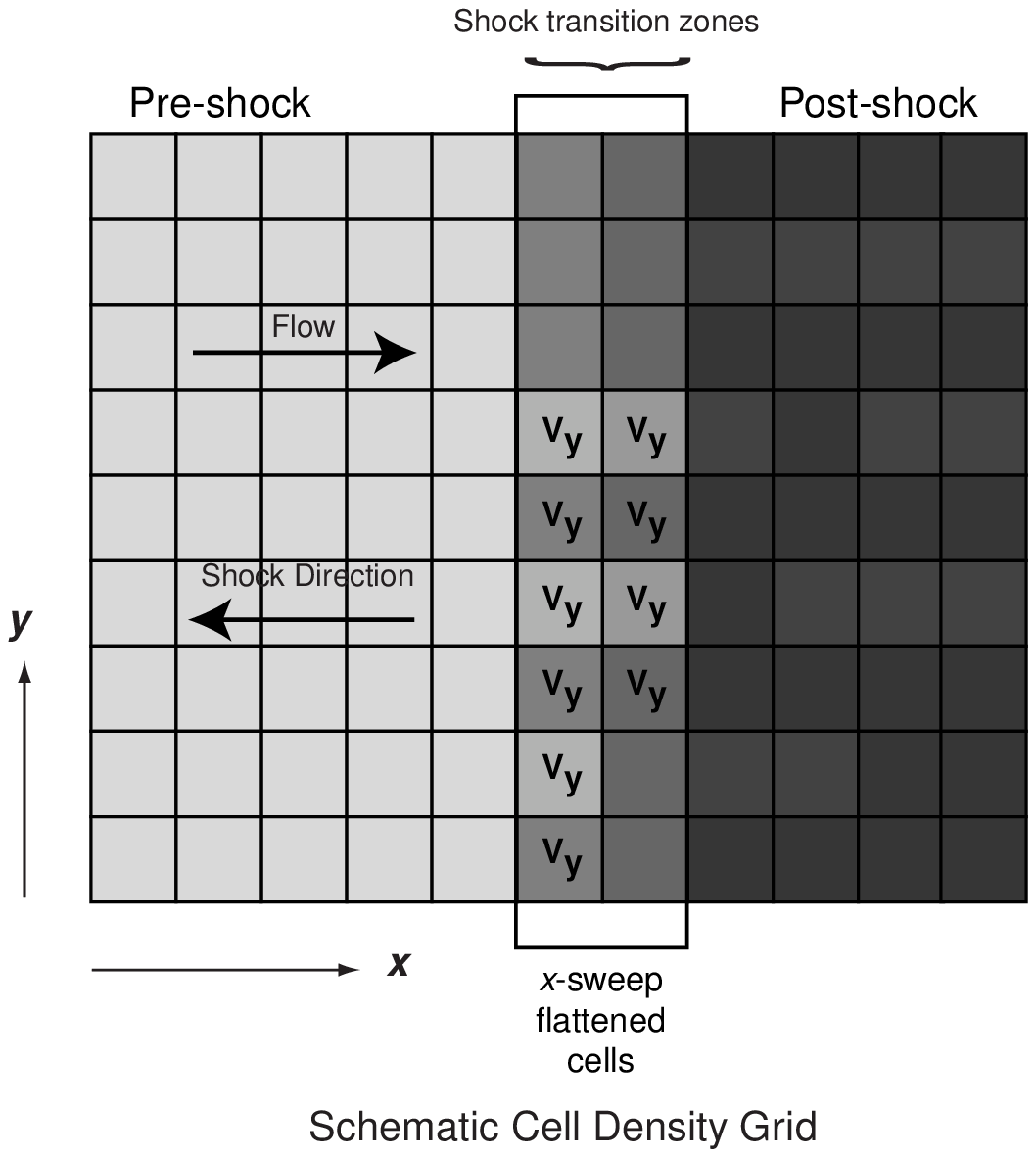}
\caption{}
\label{f:grid}
\end{figure}

The \LOF\ is constructed to use additional information from previous orthogonal sweeps (\eg\ the previous
$x-$sweep for the current $y-$sweep) as follows:
\begin{enumerate}
\item In a PPM code, the shock--capturing regions, usually
about one to three cells in extent, are clearly indicated by ``flattened'' cells. These are cells where the
piecewise parabolic interpolation is replaced by a piecewise linear interpolation, reducing to
a first order Godunov method in the shock front itself. For example, following an
$x$-sweep in a two--dimensional simulation with a shock propagating in the $x$--direction,  there is
a band of flattened cells parallel to the shock front (see Figure~\ref{f:vpstripes}). These are marked
for reference in the subsequent $y$-sweep. 
\item In the $y-$sweep, following the lagrangian advection, the code searches in only
the cells `flattened' in the prior $x-$sweep, for a specific pattern of density variations
that is indicative of striping. We require this to extend over at least five lagrangian cells and
to have at least 3 vertices, \eg\ HLHLH or LHLHL (where H is ``high'' and L is ``low'' density) .
A single vertex, or three-cell LHL or HLH pattern, occurs too frequently because of random
fluctuations and is an insufficient criterion for stripe detection. A schematic version of this procedure is given
in Figure~\ref{f:grid}. 

We take the detection threshold in the density variation, $\Delta \rho /\rho $, to be of
order
$10^{-6}$. The entire $y-$sweep row is searched for the three--vertex pattern, and the vertex cells are
marked for attention during the standard remap step, where the lagrangian cells are interpolated
back onto the Eulerian  grid.
\item Immediately subsequent to the remap step the zone averages of flagged vertex cells are reassigned.  
(A brief description of the lagrangian Riemann method
with a remap to a fixed Eulerian grid is given in the Appendix, to more clearly show where this
additional dissipation fits into a normal ppm method.)

Let $Q_i$ represent the zone average of either mass, momentum or specific energy densities ($\rho$, $\rho u$
or $E$). The reassignment of zone averages is defined in a conservative fashion by:
\begin{equation}
\begin{array}[]{l c l l}
Q_i & \rightarrow & Q_i & + \> \alpha \> (Q_{i+1} - Q_{i}) \\
Q_{i+1} & \rightarrow & Q_{i+1} & - \> \alpha \> (Q_{i+1} - Q_{i})
\end{array}
\end{equation}
Typically, $\alpha$ is no larger than $0.075$ and we have successfully used
$\alpha=0.05$.  This additional smoothing successfully damps the erroneous growing  mass
exchange and related stripe growth.
\item Cells that are flattened in the $y-$sweep as a result of shocks in that direction are marked for
potential smoothing in the subsequent $x-$sweep, and so on.
\end{enumerate}

In contrast to 1D artificial viscosity, which only smooths strong velocity gradients in
the {\em current} sweep, the \LOF\
shares information between otherwise independent sweeps, and smooths cells that
artificial viscosity does not affect. This is similar to the multi--dimensional velocity
divergence artificial viscosity from \citet{colella84b}, but the \LOF\ has the advantage in
that a much smaller number of cells are smoothed.

An important feature of this method is that in a 2D or 3D simulation, the marked cells  typically constitute
a small fraction of the cells that are associated with strong shock and that are `flattened' during the
previous sweep. The marked cells usually constitute less than 0.001\% of the total cells
in the test $320\times 64$ cell simulation since the filtering is \emph{only applied when the
up--down pattern of the striping occur in the flattened zones}. In some sweeps, no striping is
detected and consequently \emph{no} cells are smoothed.  Since additional calculations are only employed for a
limited number of cells and time--steps, the performance impact of the additional calculations for this
method is minimal, typically amounting to less than 5\% in the cells processed per second.

\section{Comparison of techniques}
\label{s:comparison}

\subsection{The test problem}
Let us now compare some of the previous techniques for handling shock--striping with the \LOF\  approach. For a
test calculation (see Figure~\ref{f:comparison}) we set up a radiative, Mach~15 flow incident from the
left upon a dense layer (a wall) at the right.  Mass flows out of the right hand grid boundary at a rate slightly
less than the mass flux of the incoming flow keeping the dense layer more or less constant in thickness during the
simulation.  The outflow velocity is sub-sonic and less than
$1$\% of the inflow velocity. The lower and upper boundary conditions are periodic. To break perfect 1D symmetry,
which \ppmlr\ handles correctly without smoothing, velocity gradients are introduced into the flow in the form of a
narrow shear layer. This is established by increasing the inflow velocity over the four cells adjacent to the
boundaries by 1, 3, 7 and 10 \% at the grid edges. The grid resolution is
$320 \times 64$ .

The physical evolution of this simulation is as follows: First, a radiative shock is reflected from the wall. Then,
as a result of post-shock cooling, the shock stalls and collapses against the cooling high density layer that forms
adjacent to the wall. The shock then re--forms, expands and collapses repeatedly. At the same time, the
Kelvin-Helmholtz instability guarantees that the shear layer is unstable and produces the characteristic pattern of
alternating vortices. The growth rate of the instability is largest in the post-shock region where the differential
Mach number across the layer is least. The growth rate of the instability is negligible in the pre-shock region
because of the hypersonic Mach number. The introduction of the shear layer to this test problem is valuable since
it enables us to judge the effect of different methods on other large velocity gradients in the flow. 

The four snapshots in Figure~\ref{f:comparison} are images of the {\em temperature} in the flow, near the point in
time at which the first shock attains its maximum distance from the wall (in the ``standard'' simulation). In each
snapshot, the periodic simulation is replicated once in the vertical direction. This has the result of showing the
unstable shear layer in the middle of each image. The grayscale of each image has been adjusted non-linearly to
enhance the salient features. In the shocked regions, white represents the coolest temperatures, black the
hottest.  In the precursor regions and the post shock cool gas, the images were scaled in a different fashion. Gray
represents a temperature of about 7500K and white is approximately 1\%  hotter. This shows the effect of  the
smoothing methods on the supersonic shear layer before it is shocked.

\subsection{The ``standard'' model}
Our first test involves using \ppmlr\ without any smoothing/anti-striping options activated; this is the `standard'
model.  Post-shock stripes of alternating higher and lower temperatures, with peak to peak variations of 10\% or
more, are evident in this simulation for about the first quarter of the post-shock flow. Downstream of this region,
additional, larger inhomogeneities are also evident. The expected instability of the shear layer is not strong,
however it spreads slightly as the interior pressure drops, and pulsations are visible.

In panel (a) of Figure~\ref{f:plots} the transverse $y$-profiles of density and pressure in
the immediate post-shock zone are shown. The mean density has almost attained its correct post-shock value
indicated by the dotted line, but there are obvious fluctuations in both density and pressure. Also note (see
specifically the cells marked 1 and 2, that the pressure and density fluctuations are anti-correlated. This is the
result of positive fluctuations in pressure pushing matter into the adjacent low pressure cells. Panel (b) of the
same figure shows the fluctuations in pressure and density several cells downstream of the shock. The pressure
fluctuations have smoothed out somewhat but the density fluctuations have grown as a result of thermal instability.

In subsequent simulations in this suite of comparisons, oblique waves are obvious in the post-shock region. These
are related to waves on the shock front. These waves are absent or appear at a low level in the standard simulation
because of the pressure perturbations associated with the stripes in the region near the shock. 

\subsection{Artificial viscosity}
The ``Lapidus artificial viscosity'' simulation was performed with the addition of code designed to implement the
artificial viscosity of \citet{lapidus67a}. This viscosity is proportional to the local velocity gradient, in the
direction of the sweep. It might be argued that this could eliminate striping, by smoothing the shock and
eliminating the irregularities that cause the stripes. Of course, an unappealing feature of this approach is that
one of the major strengths of the PPM method is that shock capturing does not normally require the use of
artificial viscosity. Some of the deficiencies of this approach are evident in the second snapshot. First, the
shock reaches its maximum extent earlier and does not expand as far as the standard shock. This is probably caused
by unphysical, additional cooling caused by incorrect densities in the post-shock zone. Second, even with a level
of artificial viscosity that causes such undesirable simulation features, post-shock striping remains. Much higher
levels of artificial viscosity would be required to reduce the striping, but it is never entirely eliminated even
with quite high values, as long as the smoothing remains aligned with the sweeps. Third, the stripes are still
strong, but have a lower spatial frequency.

\subsection{Grid jittering}
The third snapshot in Figure~\ref{f:comparison} shows the results of conducting the simulation with the
incorporation of grid--jittering. The grid was jittered in the $y$--direction only, with a speed equivalent to
seven percent of the Courant length per time step (0.042 of a grid cell with Courant number, $C = 0.6$). The
snapshot shows that this amount of grid jittering successfully counteracts striping and other tests showed that a
level of grid jittering in between five and seven percent is capable of eliminating stripes.  With the stripes
eliminated, the oblique waves, referred to above, are apparent. Nevertheless, the grid is jittered over the entire
flow, and the velocity gradients in the post-shock shear layer region are smoothed as a result. Consequently, the
shear layer does not grow rapidly (especially when compared to the local oscillation filter simulation described
below). In this case, the shear layer is also affected in the pre-shock region, with excessive diffusion resulting
in spreading of the layer and associated pressure and temperature effects. The effect of the grid-jittering in
the pre--shock region is quite unphysical. There are no Kelvin-Helmholtz vortices produced, but simply a
redistribution of the velocity resulting in a broader stream layer at the shock face than in the other
simulations. This also appears to have an effect on the degree of instability in the post--shock shear
layer.  

\subsection{Local Oscillation Filter}

The fourth snapshot shows the effect of the \LOF. This has several desirable features. The post--shock striping is
absent; the shear layer starts to become unstable almost immediately following the shock and although there are
initially only less than six pixels across the shear layer the detail is excellent. At the contact discontinuity
between the incoming flow and the dense layer adjacent to the wall, there is more detail in the flow. With the
stripes absent, the oblique wave pattern noted above, is quite apparent.  The pre--shock region and shear
layer are unaffected. 

The amplitude of this simulation is very close to the standard model and the grid--jittered model. The general
features of the simulations are similar to those of the grid--jittered model, in general, except that the 
resolution appears better.  The fraction of cells that have been smoothed in this simulation averages to about
$6\times 10^{-4}$ of the total cells, over the total of approximately 2300 $x-y$ iterations.  Additionally, in
about 46\% of the sweeps, no smoothing took place at all, and frequently less than 10\% of the cells adjacent to
the shock front region are smoothed in a given sweep.  This means that most of the cells are untouched by the
smoothing and the simulation retains the maximum resolution allowed by the ppmlr method.  The decrease in code
performance resulting from additional testing and smoothing in this simulation is less than 4\% overall.  This
small simulation has a high I/O overhead however and the impact on larger simulations is expected to be less.

\section{A More Severe Test}

With only 10\% shear velocities, the grid--jittering and the \LOF\ perform similarly well.  However, with stronger
shear layers, the excessive diffusion introduced by the jittering technique is unsatisfactory.  The Local
Oscillation Filter is designed so as not to affect shear layers, or even large steps in velocity, but to focus upon
the characteristic up--down noise of the striping error.

In order to see at what point grid jittering becomes a problem, and whether the \LOF\ can cope with the same
conditions, we performed a simulation with an increased shear gradient, maintaining the same smoothing as in the
first simulation - that is sufficient jittering and \LOF\ smoothing to just remove the stripes.  The shear was
increased to 1, 5, 20, and 50\% velocity increase at the grid boundary.  This is more extreme, but still less than
a factor of unity compared to the bulk velocity, and thus is not an unreasonable test case.

The results of the standard, jittered and \LOF\ models are shown in Figure \ref{f:bigshear}, at a model step
corresponding to 40\% of the peak shock amplitude in the $x$--direction.  The standard model shows the usual
striping, and some  shear instability in the post--striping region.  The Local Oscillation Filter model has no
stripes, the same shock front location, some oblique waves, and the shear instability is well developed in the
post--shock region.  The stream has also started to dig a hole in the dense wall layer, depositing energy there. 
The grid--jittering model, however, is different.  The shear layer in the precursor region has been diffused so
much by the jittering that the wider stream of gas has a significantly exerts a different pressure over a large
fraction of the shock front, strongly deforming the shock front in the region of the shear layer impact.  The post
shock shear layer is completely disrupted and less energy is deposited into the dense layer, probably accounting
for the increased amplitude of the shock.

We conclude that in the case of a strong shear layer parallel to the grid, jittering can  cause quite major
errors.  By operating as a localized spatial filter, the Local Oscillation Filter avoids this particular problem,
having no effect on the regions outside the shock front location.

\section{Conclusions}

The work we have described here represents the first step in a project aimed at the accurate simulation of
radiatively cooling flows, particularly those which contain embedded shocks. As we have seen, the inclusion of
cooling is not simply a matter of adding the right terms to the energy equation. One also needs to consider
numerical instabilities that may be enhanced by thermal instabilities. In this paper we have investigated a classic
case, that of post-shock striping that occurs when shock fronts are parallel to a coordinate grid. This is not a
contrived situation and occurs, for example, near the axis of a supersonic jet simulation. Our local oscillation
filter solution is simple and the tests that we have presented show it to be effective. It has the advantage that
it does not affect other regions of the flow, nor does it significantly affect the overall performance of the code.

\subsection*{Appendix: The Lagrangian -- Eulerian PPM method with Additional Remapping Dissipation}
\label{a:ppm}

As is common in  methods based on the PPM algorithm, \emph{VH1} and \emph{ppmlr} use a Lagrangian time step,
where the fluxes at the cell boundaries are evaluated from the solution of the local Riemann problem. After
moving the boundaries,
$\xi_{j+1/2}$, which are contact discontinuities, the solution is remapped to the original  Eulerian grid,
$\xi_{i+1/2}$. (Note that we use a subscript $i$ for the Eulerian grid and a $j$ for the Lagrangian grid.  
Proper selection of the Courant number
$C$, prevents the boundaries from moving more than an Eulerian grid cell in a single time step,
keeping the remapping process simple and accurate.

Solving the Riemann problem at the Lagrangian cell boundary, $\xi_{j+1/2}$, gives the velocity at the
contact discontinuity $u_*$.  After parabolic interpolation over the region within the domain of
dependence of the solution at $\xi_{j+1/2},t^{n+1}$, the left and right states $W_L$ and $W_R$ of the
primitive variables $\rho$, $u$, and $p$ are known. The Riemann problem is then solved for the region between
the two outer waves -- the so-called star region.  The velocity and pressure, $u_*$ and $p_*$, in this
central region are constant, and $u_*$, velocity of the  contact discontinuity, is given  by
\begin{equation}
u_* = \frac{1}{2} (u_L + u_R) + \frac{1}{2} [f_R(p*) - f_L(p*)] ,
\end{equation}
where $u_L$ and  $u_R$ are the initial left and right velocities, and $f_R(p*)$ and $f_L(p*)$ are functions
determined by the nature of the outer waves in the Riemann solution; these are either rarefaction or shocks. 
For a clear and detailed description of how $f_R(p*)$ and $f_L(p*)$ are determined, see \citet{toro99a}. 
Following the solution to the Riemann problem on each cell boundary, the coordinates of the cell boundary
$\xi_{j+1/2}$ are moved with the velocities $u_*$ over the timestep $\uDelta t$,
\begin{equation}
\xi_{j+1/2}^{n+1} = \xi_{j+1/2}^{n} + u_{*,j+1/2} \uDelta t.
\end{equation}
The cell boundary follows the contact discontinuity in the Riemann problem, so that at the end of the step
each cell is naturally constrained to contain the same mass as at the beginning of the time step.

Using parabolic interpolation in the volume coordinate (as described in CW84), the hydrodynamic grid is
remapped back to the original Eulerian grid in the following way. The volume $\uDelta V$ of the overlapping
regions between the updated Lagrangian grid, $\xi_j$  and the original Eulerian grid $\xi_i$ , is determined:
\begin{equation}
\uDelta V_{i+1/2} = \xi_{j+1/2} - \xi_{i+1/2}.
\end{equation}
Then, effective mass, momentum, and energy fluxes, $F_{i+1/2}$ between the Eulerian cells are evaluated from:
\begin{equation}
F_{i+1/2} = \uDelta V_{i+1/2} \langle q_{i+1/2} \rangle,
\end{equation}
where the $\langle q_{i+1/2} \rangle$ represent the averages of the relevant dynamic quantities ($q=\rho$,
$\rho u$ or $E$) over the volume
$\uDelta V_{i+1/2}$. The remapping fluxes are used to  redistribute the Lagrangian gridded variables onto 
the Eulerian grid, and the pressure is updated at the end from $\rho$, 
the velocity, $u$ and the total specific energy, $E$. Note that the fluxes, $F_{i+1/2}$ are positive if the
Lagrangian boundary moves to the right, because of the convention in defining $\uDelta V_{i+1/2}$.  The
remapping procedure maintains the conservative nature of the Lagrangian method.

The Local Oscillation Filter transports a small amount of mass, momentum and energy in the opposite
direction in the flagged cells. Thus, immediately following the remap step, the Eulerian zone average, $Q_i$
of a given quantity, $q_i$ is modified by:
\begin{equation}
\begin{array}[]{r c l l}
Q_i & \rightarrow & Q_i & + \> \alpha \> (Q_{i+1} - Q_{i}) \\
Q_{i+1} & \rightarrow & Q_{i+1} & - \> \alpha \> (Q_{i+1} - Q_{i})
\end{array}
\label{e:flux_ansatz}
\end{equation}
where the coefficient $\alpha \la 0.075$. Hence, an excess (deficit) of $Q_{i+1}$ over $Q_i$ is reduced
(increased) by a small amount and in a conservative fashion. The zone average is modified only if both the
cell has been flattened during a previous orthogonal sweep and also if the pattern of variations in density
satisfies the ``HLHLH'' criteria. This ansatz amounts to a modification of the remapping fluxes
according to
\begin{equation}
F_{i+1/2} \rightarrow F_{i+1/2} - \alpha (Q_{i+1} - Q_i)
\end{equation}
bearing in mind that the $Q_i$ have been obtained from the previously unmodified fluxes. This reassignment
of zone averages critically damps the instability. The value of $\alpha \approx 0.075$ was arrived at
through experimentation.

In summary, the modification of fluxes is implemented in the following way. Cells that are flattened in
one sweep are marked (using an integer array). When an HLHLH or LHLHL density pattern is detected in marked
cells in one of the subsequent orthogonal sweeps, the remapped zone averages of all variables into and
out of the vertex cells in the pattern are modified using equation~(\ref{e:flux_ansatz}). For example in our
test problem, flattened cells are detected in the
$x$--sweep. When the appropriate density variation pattern is picked up in the ensuing $y$--sweep, the
zone averages in the $y$-direction are modified in such a way as to partially reverse the numerical fluxes into
and out of the vertex cells in the pattern. A critical point in our LOF algorithm is that this
effective dissipation is only applied in restricted  regions but not globally.  This prevents
unnecessary dissipation in regions such as shear layers;  regions such as these are affected by all of the
other algorithms that we have discussed.

\newpage

\begin{center}
\bf Figure Captions
\end{center}

\noindent{\bf Figure 1:} A portion of a typical 2D strong radiative shock model using a directionally split sweep
code.  Striping can occur at the stagnation point of bowshocks (a), but may be absent altogether when the shock is
oblique to the grid (b).  Flat portions of a nearly stationary shock (c) are particularly prone to this type of
error.  All of these situations can occur simultaneously in a given simulation.

\noindent{\bf Figure 2:} A small portion of a Mach 15 shock simulation in two dimensions, with uniform, one
dimensional initial conditions.  Small grid errors in VH-1 stimulate the stripe instability. In the \ppmlr\ code the
grid errors are removed and the system remains pseudo one-dimensional indefinitely.  However, small perturbations
still cause strong striping in \ppmlr\ unless further steps are taken to control the instability.

\noindent{\bf Figure 3:} A schematic representation of the \LOF\ smoothing, showing the grid density cells.  In the
$x-$sweep, cells in the shock transition zone use a lower order interpolation scheme, and are said to be
'flattened'.  In the $y-$sweep, cells in the $x-$flattened cells are searched for three high-low-high or
low-high-low density 'vertices' in a row (requiring 5 high--low density points in a row).  In the diagram, only the
cells marked $V_y$ are density vertices by this definition in the $y-$sweep.  Of these cells, only the cells in the
left column will be smoothed by the \LOF\ because there are more than 5 cells marked in a row.  The shorter run of
4 vertices in the right column will not be smoothed. In practice, the vertices in the right column will not occur,
because that gas will have already been smoothed when passing through the left column of the grid.

\noindent{\bf Figure 4:} Snapshots of temperature, at the same instant of time, of four different wall-shock
simulations comparing different approaches to the elimination of post--shock striping.  In each case, a Mach 15
supersonic inflow enters from the left, producing a reflected shock from a dense wall layer on the right.  A shear
layer with increased velocity in steps of 1\%, 3\%, 7\% and 10\% at the edges of each grid slightly modifies the
planar nature of the shock front and induces particularly strong striping.  In the figures, the periodic nature of
the boundaries in the $y$--direction is shown by repeating the grid, showing the shear layer as a stream in the
middle of the simulation.  The gray scale is a non--linear scaling applied to the logarithm of the temperature, to
best show the small ($\le 10$\%) fluctuations on top of the intrinsic 10-100 fold variations in temperature caused
by the shock.  In the precursor region the contrast enhancement is more extreme to show variations $<<1$\%,
revealing small temperature errors in the grid--jittering model which are absent in the other simulations.

\noindent{\bf Figure 5:} Plots of the density and pressure in the immediate
post shock zone (panel (a)) and a few zones downstream of the shock (panel (b)). Panel (a) shows significant
fluctuations in density and pressure as a result of the instability. The density and pressure are anti--correlated
as a result of the perturbations driving mass from the high pressure zones into the low pressure zones. In panel
(b) the pressure fluctuations have diminished but the density fluctuations have survived as a result of the thermal
instability.

\begin{figure}[tbp]
\epsscale{0.80}
\plotone{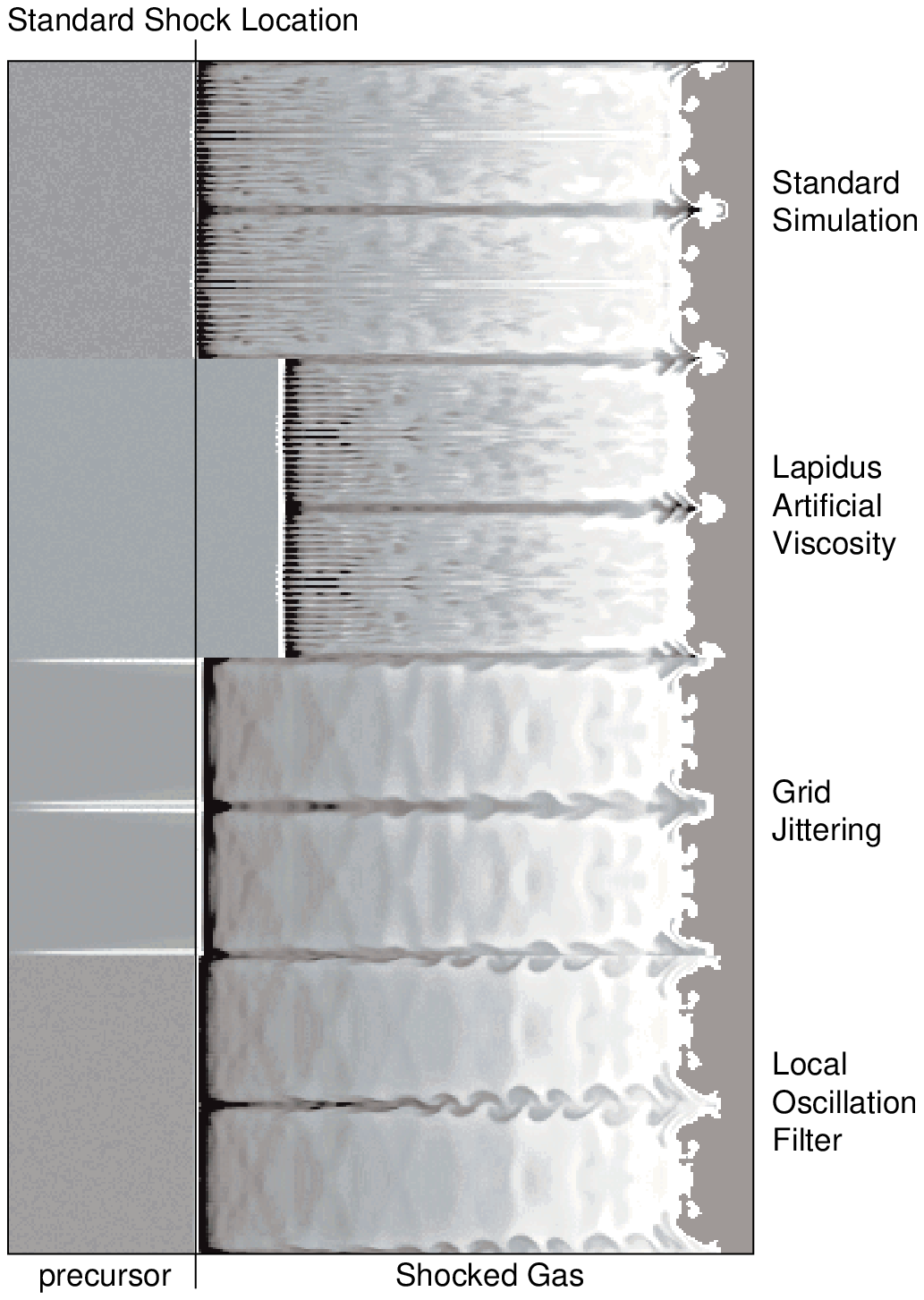}
\caption{}
\label{f:comparison}
\end{figure}

\begin{figure}[tbp]
\plotone{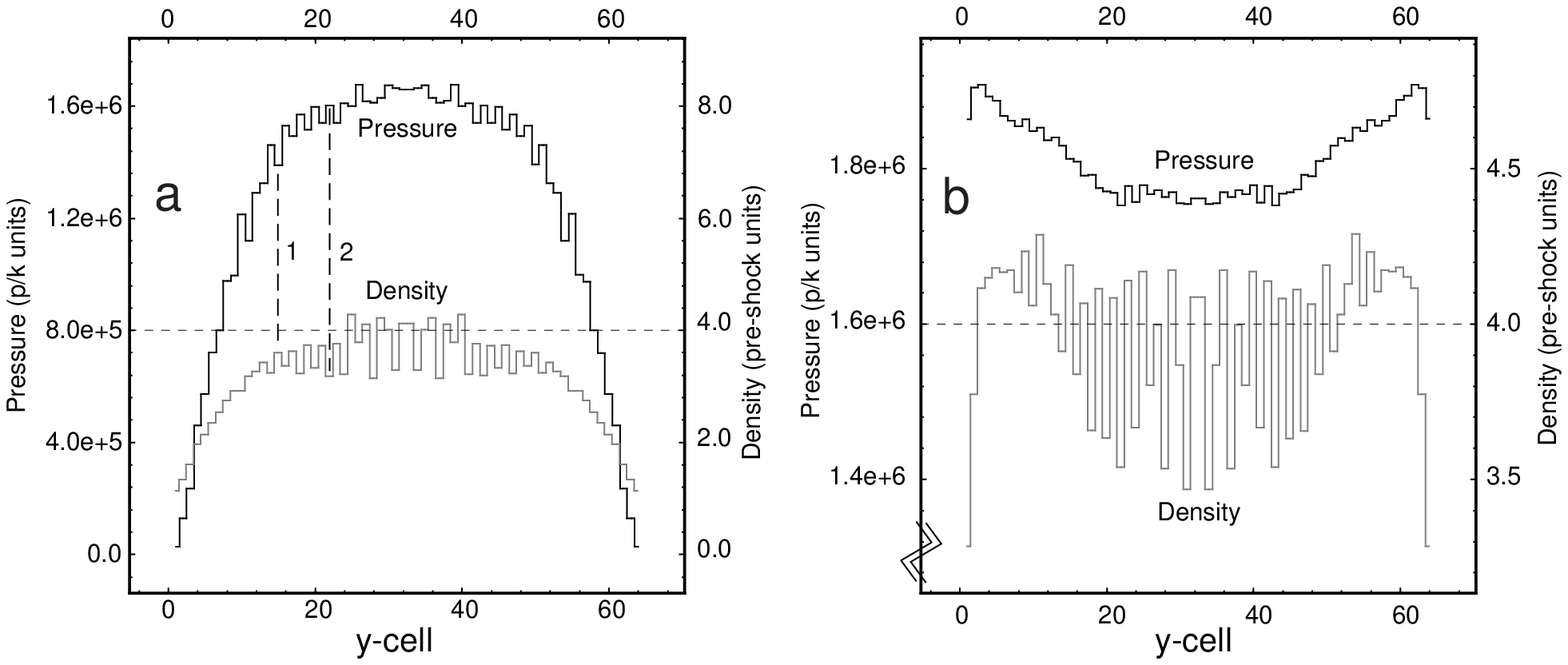}
\caption{}
\label{f:plots}
\end{figure}

\begin{figure}[tbp]
\plotone{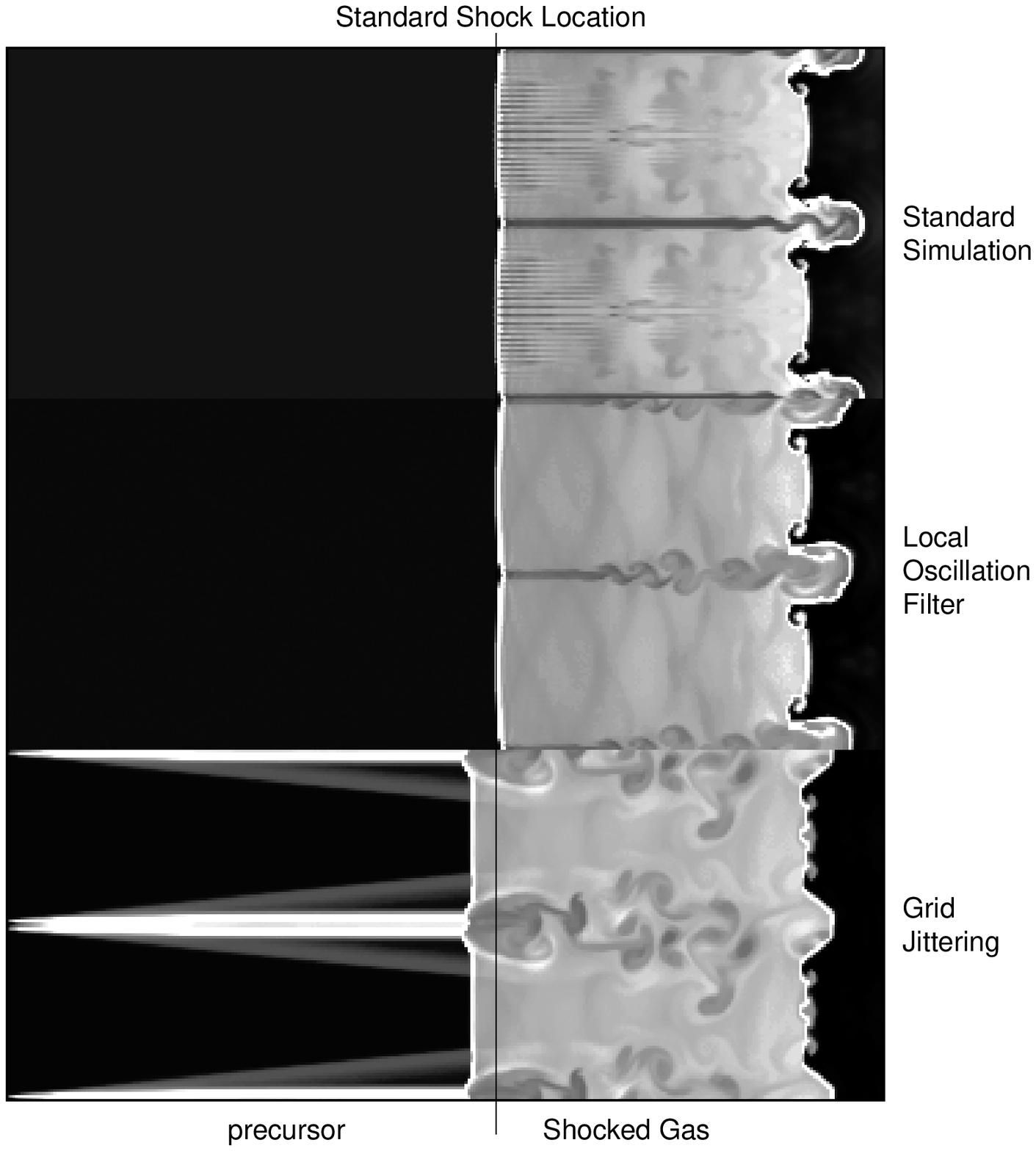}
\caption{}
\label{f:bigshear}
\end{figure}

\noindent{\bf Figure 6:} Snapshots of temperature, at the same instant of time, of three different simulations,
similar to figure \ref{f:comparison}, but with an increased shear profile of  1\%, 5\%, 20\% and 50\%  over the
velocity of the incoming stream.  The pre-shock jittering spreads the shear layer, substantially modifying the shock
front, disrupting the post shock stream completely, resulting in a different shock amplitude and appearance. The
\LOF\ simulation eliminates the standard model's stripes without affecting the shear layer at all.

\newpage

\end{document}